\documentclass[twocolumn,showpacs,preprintnumbers,superscriptaddress]{revtex4}
\usepackage{graphicx}
\usepackage{dcolumn}
\usepackage{bm}
\usepackage{color}
\usepackage{amsmath}
\usepackage{amssymb}
\usepackage{amsfonts}

\begin{document}

\title{Witnessing topological Weyl semimetal phase in a minimal circuit-QED lattice}

\author{Feng Mei}
\email{tianfengmei@gmail.com}
\affiliation{TianQin Research Center and School of Physics and Astronomy,
Sun Yat-Sen University, SYSU Zhuhai Campus, Zhuhai 519082, China}

\author{Zheng-Yuan Xue}
\affiliation{Guangdong Provincial Key Laboratory of Quantum Engineering and Quantum Materials, SPTE, South China Normal University, Guangzhou 510006, China}

\author{Dan-Wei Zhang}
\affiliation{Guangdong Provincial Key Laboratory of Quantum Engineering and Quantum Materials, SPTE, South China Normal University, Guangzhou 510006, China}

\author{Lin Tian}
\email{ltian@ucmerced.edu}
\affiliation{School of Natural Sciences, University of California, Merced, California 95343, USA}

\author{Chaohong Lee}
\email{lichaoh2@mail.sysu.edu.cn}
\affiliation{TianQin Research Center and School of Physics and Astronomy,
Sun Yat-Sen University, SYSU Zhuhai Campus, Zhuhai 519082, China}

\affiliation{State Key Laboratory of Optoelectronic Materials and Technologies,
Sun Yat-Sen University, SYSU Guangzhou Campus, Guangzhou 510275, China}

\author{Shi-Liang Zhu}
\email{slzhunju@163.com}
\affiliation{National Laboratory of Solid
State Microstructures, Department of Physics, Nanjing University,
Nanjing, China}

\affiliation{Guangdong Provincial Key Laboratory of Quantum
Engineering and Quantum Materials, SPTE, South China Normal
University, Guangzhou 510006, China}

\affiliation{Synergetic Innovation Center of Quantum Information and Quantum Physics,
University of Science and Technology of China, Hefei 230026, China}

\date{\today}

\begin{abstract}
 We present a feasible protocol to mimic topological Weyl semimetal phase in a  small one-dimensional
 circuit-QED lattice. By modulating the photon hopping rates and on-site photon frequencies in parametric spaces,
 we demonstrate that the momentum space of this one-dimensional lattice model can be artificially mapped to three
 dimensions accompanied by the emergence of topological Weyl semimetal phase. Furthermore, via a lattice-based cavity
 input-output process, we show that all the essential topological features of Weyl semimetal phase,
 including the topological charge associated with each Weyl point and the open Fermi arcs, can be unambiguously
 detected in a circuit with four dissipative resonators by measuring the reflection spectra.
 These remarkable features may open a new prospect in using well-controlled small quantum lattices to mimic and study topological phases.
\end{abstract}

\pacs{85.25.Am, 85.25.Cp, 03.65.Vf}

\maketitle

\section{Introduction}
Circuit quantum electrodynamics (QED) has achieved great
experimental progresses in the past decade, including
demonstrating  surface-code based error correction
\cite{Surfacecode,KellyNature2015}. It turns this system into one
of the leading quantum platforms for studying scalable quantum
computation \cite{Devoret2013, NoriQOQC}. The experimental
progresses also endow circuit QED system with high coherence
superconducting qubits and microwave resonators
\cite{Martinis2013,Schoelkopf2015}, tunable system parameters and
straightforward connectivity
\cite{Chen2014,Houck2013,Wallraff,Blais}. These advantages further
push this system towards quantum simulation
\cite{CircuitQS,NoriQS,circuitkoch,LeHur}. In particular, circuit
QED lattice framed by coupling microwave resonators and
superconducting qubits has been widely pursued as analogue quantum
simulators to study various many-body physics
\cite{PolaritonQS,Rosario,Polaron1,Polaron2,Solano,DQS1,DQS2}.
Experimentally, digital quantum simulation of quantum spin models
has recently been realized based on superconducting circuit
\cite{DQS1,DQS2}.

The recent advent of topological photonics has opened a door for
studying topologically protected states with  photons
\cite{Haldane,LuRev,Marin}. In particular, photonic topological
insulator states nowadays have been extensively studied in
different photonic systems
\cite{Alexander,Chan,Segev,Hafezi,Carusotto,Hur,Chong,Bardyn,Zhou}.
In the context of circuit QED, the methods to produce photonic
synthetic gauge fields have been investigated in a two-dimensional
coupled microwave resonator lattice \cite{Girvin,Ripoll,Yang}.
Based on engineering photonic magnetic fields and photonic
interactions, integer and fractional photonic topological
insulator states were also studied \cite{Hu,Taylor,Kapit}. Another
route realizing photonic topological phase is based on dimension
reduction \cite{Kraus,Mei2015,Chen,MeiCA,Zhu}. It is found that
the topological features of two-dimensional (2D) photonic topological
insulators could be observed in a one-dimensional (1D) system
\cite{Kraus,Mei2015}, where one parameter that can be tuned
periodically has been introduced as an artificial dimension in
addition to the real spatial dimension.

On the other hand, topological Weyl semimetal now has
attracted significant attentions because its low energy
excitations are Weyl fermions
\cite{Wan,Burkov,Fang,Weng,Hasan,Alexey,Zhang}. This gapless
semimetal phase is a new topological state of matter, extending
topological classification to go beyond gapped topological
insulators and opening a new era for condensed matter physics.
Specifically, the conduction and valence bands of topological Weyl
semimetal touch nontrivially at pairs of Weyl points \cite{FDT}.
Both the Weyl fermions in the bulk and the Fermi arcs on the
surfaces are interesting in topological Weyl semimetal state,
which leads to various exotic physical properties
\cite{Nielsen,Son,Hosur}. The experimental realization of
topological Weyl semimetal was first reported in TaAs materials,
where both the Weyl cones and the Fermi arcs were observed
\cite{Hasanexp,Dingexp}. Weyl points have also been observed in
photonic crystal system \cite{LuWP}, and have been theoretically
studied in cold atoms \cite{WSMCA} and acoustic systems
\cite{WSMAS}. However, the question of directly measuring the
topological charge associated with each Weyl point remains
unsolved either in solid-state materials or in artificial lattice
systems.

In this paper, we propose a feasible protocol to mimic topological
Weyl semimetal phase with a 1D circuit-QED
lattice. In this lattice, the photon hopping between nearest
neighbour resonators and the frequency shift of the resonators can
be tuned by externally control circuit with the state-of-art
circuit-QED technology, which allows us to parameterize these two
parameters and employ them as two artificial momentum dimensions.
Combining these two synthetic momentums with the momentum in the
real space, we construct a synthetic three-dimensional (3D)
momentum space, and we find that four Weyl points emerge in the
first Brillouin zone. The resulted gapless phase is further
demonstrated as a topological Weyl semimetal phase by numerically
analyzing the topological charges of the Weyl points and the Femri
arc surface states. More importantly, with a lattice-based cavity
input-output process, we show that all the essential features of
topological Weyl semimetal can be unambiguously measured from four
coupled resonators with dissipation. In particular, we design a
simple strategy to measure the topological charge associated with
each Weyl point, which has not been shown in the literature
before. Such remarkable results render us a  minimal and realistic
platform to observing topological Weyl semimetal physics and make
our theory attractive to the experimentalists in the circuit-QED
community.

The paper is organized as follows. In section II, the 1D
circuit-QED lattice for simulating the Weyl semimetal and its
circuit parameters  are presented. The mapping of the controllable
parameters to the 3D momentum space and the emergence of gapless
Weyl semimetal phase are studied in section III. The essential
features of our system associated with the topological Weyl
semimetal phase explicitly given in section IV. Finally, the
measurement of the topological features with a small lattice is
analyzed in Section V. The paper is summarized in Section VI.

\begin{figure}[h]
\includegraphics[width=8.5cm,height=2.5cm,clip]{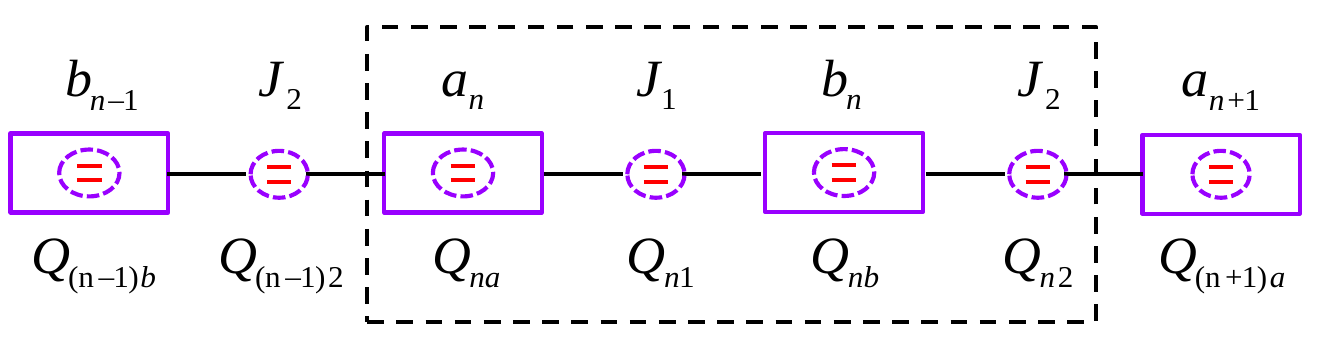}
\caption{Schematic setup of the one-dimensional circuit QED
lattice. The resonators are  labelled as $a_{n}$ and $b_{n}$ with
alternative coupling strength $J_{1,2}$. Neighboring resonators
are connected by qubits $Q_{n1}$ and $Q_{n2}$, and each resonator
is coupled to a qubit embedded inside the resonator. In this
circuit, both the photon hopping rates and on-site photon
frequencies have an alternative configuration. The dashed box
defines the unit cell of this lattice. }
\end{figure}

\section{One dimensional circuit-QED lattice}
We consider a 1D array of superconducting transmission line
resonators and Xmon qubits alternatively  connected to each other.
The crossed-shape Xmon qubit has recently attracted much attention
as it can be used to construct scalable quantum circuits
\cite{Martinis2013,Chen2014}. The Xmon qubit can be connected to
different circuit elements simutaneously, including resonators or
external qubit control lines. Such feature allows us to build the
circuit-QED lattice shown in Fig.~1. In this lattice, each unit
cell contains two resonators, labelled by $a_{n}$ and $b_{n}$,
respectively. The two resonators $a_n$ and $b_n$ in the $nth$ unit
cell are both coupled to the Xmon qubit $Q_{n1}$, and the two
resonators $b_n$ and $a_{n+1}$ belonging to two adjacent unit
cells are both coupled to the Xmon qubit $Q_{n2}$. Meanwhile,
another Xmon qubit is embedded inside each resonator to provide
additional control over the frequency shift of the resonator
modes. Specifically, we assume the two resonators $a_n$ and $b_n$
in the same unit cell are coupled to the Xmon qubits $Q_{na}$ and
$Q_{nb}$, respectively. The total Hamiltonian of this system can
be written as
\begin{eqnarray}
H_s &=& \sum_{n}\sum_{x=1,2,a,b}\frac{\omega_x}{2}\tau_{nx}^{z}+\omega_c(a^{\dag}_na_n+b^{\dag}_nb_n) \nonumber \\
&&+g_1\tau_{n1}^{+}(a_n+b_n)+g_2\tau_{n2}^{+}(a_{n+1}+b_n)  \nonumber \\
&&+g_a\tau_{na}^{+}a_n+g_b\tau_{nb}^{+}b_n+H.c.,
\end{eqnarray}
where $\omega_{x}$ and $\omega_c$ are the frequencies of the
superconducting Xmon qubits (with $\tau_{nx}$ operators)  and the
transmission line resonators , respectively, $g_{1}$ ($g_2$) is
the coupling strength between the Xmon qubit $Q_{n1}$ ($Q_{n2}$)
and the resonators $a_n$ and $b_n$ ($b_n$ and $a_{n+1}$), and
$g_{a}$ ($g_b$) is the coupling strength between the resonator
$a_n$ ($b_n$) and the Xmon qubit $Q_{na}$ ($Q_{nb}$).

The system is operated in the dispersive regime, where the qubit
frequencies are far off resonance from that of  the resonators
\cite{circuitQEDtheory}. In this regime, the qubit-resonator
couplings result in effective photon hopping between adjacent
resonator modes with the hopping rates controllable via the qubit
frequencies. Such coupling also generates frequency shift of the
resonator modes, which can also be manipulated by adjusting the
qubit frequencies. Here we assume all  Xmon qubits are prepared in
their ground state so that the resonators and the Xmon qubits are
effectively decoupled except for the qubit-mediated photon hopping
and qubit-induced resonator frequency shift. As we will show
later, external microwave driving on the resonators will be
employed for the detection of the topological effects. For this
purpose, we choose to discuss this system in the rotating frame
with respect to the external driven frequency $\omega_d$ and in
the interaction picture with respect to the qubit frequencies
$\omega_{x}$. The total effective Hamiltonian in the dispersive
regime then has the following form
\begin{align}
H&=\sum_{n}(J_1a^{\dag}_nb_n+J_2a^{\dag}_{n}b_{n-1}+H.c.) \nonumber \\
&+(\Delta_c-\chi-\frac{g^2_a}{\Delta_a})a^{\dag}_na_n+(\Delta_c-\chi-\frac{g^2_b}{\Delta_b})b^{\dag}_nb_n,
\label{eq:Heff}
\end{align}
where the qubit-assisted photon hopping rates between resonators
are $J_{1,2}=-g^2_{1,2}/\Delta_{1,2}$ with the cavity (qubit)
detuning $\Delta_c=\omega_c-\omega_d$
($\Delta_{x}=\omega_{x}-\omega_d$), and the resonator frequency
shift induced by the inline Xmon qubits is
$\chi=-g^2_1/\Delta_1-g^2_2/\Delta_2$.

The effective parameters in the above 1D model are fully
controllable, where the photon hopping rates and the resonator
frequency  shift could be directly tuned by changing the qubit
detuning $\Delta_{x}$ $(x=1,2,a,b)$ via external control circuits.
In experiment, this is achieved by adjusting the qubit frequency
$\omega_x$. Note that the cross-shaped Xmon qubits can be
connected to multiple ports simutaneously. In addition to the
ports that connect to the resonators, external current line can be
coupled to the qubits to generate flux bias and control the qubit
frequency. The Xmon-qubit-based circuits in Fig. 1 have already
been studied experimentally and the control of the qubit frequency
is a standard technology~\cite{Surfacecode,
KellyNature2015,Chen2014, Martinis2014}. With a qubit-resonator
coupling strength on the order of $300$ MHz and a detuning on the
order of $2$ GHz, the photon hopping rates are in the range of
$50$ MHz, far exceeding the scale of the resonator bandwidth and
the qubit decoherence rate. The topological effects studied in
this work can hence be probed in practical circuits. Such circuits
have also been studied in other parameter regimes, where quantum
phase transitions for cavity polaritons or spin particles can be
observed~\cite{GarciaRipollPRL2014, SeoPRB2015}. We also want to
mention that tunable photon hopping between resonators can also be
generated with other circuit setups, such as a tunable SQUID
loop~\cite{PeropadrePRB2013,TianNJP} or a tunable
capacitor~\cite{AverinPRL2003}.

\section{Three-dimensional (3D) Weyl semimetal phase}
By varying the the photon hopping rates and the shifts of the
resonator frequency, we can construct a series of  lattice
Hamiltonians. Specifically, we focus on lattice Hamiltonians
confined to the following parameter conditions:
 \begin{equation}
 \begin{aligned}
 J_1 &= J[1-\cos(\theta_1)],\\
 J_2 &= J[1+\cos(\theta_1)],\\
 g^2_a/\Delta_a &= J_e[1-\cos(\theta_2)], \\
 g^2_b/\Delta_b &= J_e[1+\cos(\theta_2)],
 \end{aligned}\label{eq:para}
 \end{equation}
where the parameters $\theta_1$ and $\theta_2$ can be changed from
$0$ to $2\pi$. By combining the lattice momentum $k_x$ associated
with the 1D circuit, $\theta_1$, and $\theta_2$, we thus construct
an effective 3D artificial momentum space with
$\textbf{k}=(k_x,\theta_1,\theta_2)$. Below we study the
underlying topological features in the synthetic first Brillouin
zone: \{$k_x\in(0,\pi],\theta_1\in(0,2\pi],\theta_2\in(0,2\pi]$\}.

\begin{figure}[h]
\includegraphics[width=8.5cm,height=4cm]{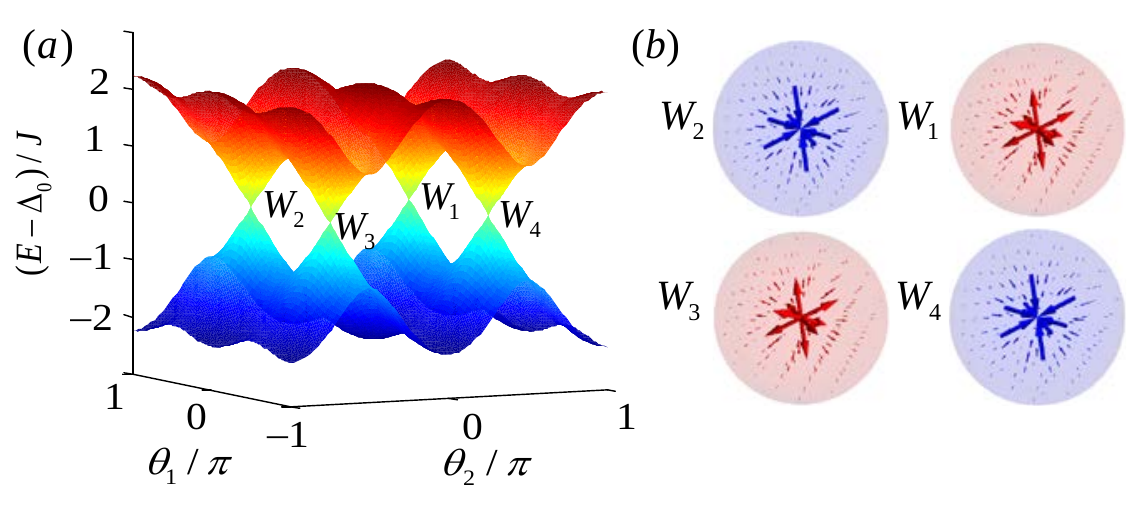}
\caption{(a) The bulk energy spectrum in the $k_x=\pi/2$ plane
with four Weyl points. (b) The Berry curvature distribution near
the Weyl points. The field lines of the curvatures are indicated
by the arrowed lines. Red: $C=1$; and blue: $C=-1$. We choose
$J_e=J$, where $J$ is used as the energy unit.}
\end{figure}

Substituting the parameters in Eq.~(\ref{eq:para}) into the
Hamiltonian (\ref{eq:Heff}) and applying the Fourier
transformation  in momentum basis, we obtain the Hamiltonian in
the momentum space with
$H=\sum_{\textbf{k}}c^{+}_{\textbf{k}}h(\textbf{k})c_{\textbf{k}}$,
where $c_\textbf{k}=(a_\textbf{k}, b_\textbf{k})^T$ is a vector
composed of the momentum-space operators $a_\textbf{k}$ and
$b_\textbf{k}$. The momentum-space matrix is
\begin{equation}
\label{hk}
h(\textbf{k})=\Delta_0+h_x\sigma_x+h_y\sigma_y+h_z\sigma_z,
\end{equation}
where $\Delta_0 = \Delta_c-2J-J_e$, $h_x =2J\cos(k_x)$, $h_y
=2J\cos(\theta_1)\sin(k_x)$, $h_z =J_e\cos(\theta_2)$, and
$\sigma_{x,y,z}$ are  the Pauli matrices spanned by the two
momentum components $a_{\textbf{k}}$ and $b_{\textbf{k}}$. With
this equation, we find that there are four band-touching points in
the synthetic 3D first Brillouin zone with
$\textbf{k}_\textbf{w}=(\frac{\pi}{2},\pm\frac{\pi}{2},\pm\frac{\pi}{2})$,
respectively. In the following, we will show that these four
band-touching points are Weyl points and the low-energy spectrum
near the band-touching points describes a gapless Weyl semimetal
state. The energy spectrum corresponding to Eq. (\ref{hk}) is
plotted in Fig. 2(a). This result indicates that the low-energy
spectrum around the band-touching points is gapless. This gapless
phase has a resemblance to a 3D Weyl semimetal phase with
$\theta_1$ and $\theta_2$ playing the role of the momenta $k_y$
and $k_z$.

For this purpose, we first expand the Hamiltonian $h(\textbf{k})$
around the four band-touching points $\textbf{k}_\textbf{w}$.  The
corresponding low-energy Hamiltonians can be derived as
\begin{equation}
\label{hw}
H_w(\textbf{q})=-2Jq_x\sigma_x\mp 2Jq_y\sigma_y\mp J_eq_z\sigma_z,
\end{equation}
where
$\textbf{q}=(q_x,q_y,q_z)=\textbf{k}_\textbf{l}-\textbf{k}_\textbf{w}$
and $\textbf{k}_\textbf{l}$ is the synthetic momentum  near the
touching points. One can find that the above Hamiltonians are
exactly the Weyl Hamiltonians and the band-touching points are the
Weyl points. In general, the Weyl Hamiltonian can be expressed as
$H_w=\sum_{ij}v_{ij}q_i\sigma_j$, which describes a massless Dirac
fermion with a fixed chirality defined as
$Ch=\text{sign}(\text{Det}[v_{ij}])$. For our system, the
chirality for the two Weyl points
$W_{1}=(\frac{\pi}{2},\frac{\pi}{2}, \frac{\pi}{2})$ and
$W_{3}=(\frac{\pi}{2},-\frac{\pi}{2}, -\frac{\pi}{2})$ is $1$; and
the chirality becomes $-1$ for the two Weyl points
$W_{2}=(\frac{\pi}{2},\frac{\pi}{2}, -\frac{\pi}{2})$ and
$W_{4}=(\frac{\pi}{2},-\frac{\pi}{2}, \frac{\pi}{2})$.

\section{Essential topological features}
In this section, we will demonstrate in detail that the above
gapless state is in a topological Weyl semimetal phase. Firstly,
we analyze the topological stability of the Weyl points by
calculating their Berry curvature in the 3D momentum space. We
find that each Weyl point can be seen as a magnetic monopole with
the monopole charge expressed by a Chern number. This is a key
proof that connects the Weyl semimetal to a topological phase.
Secondly, we numerically calculate the edge-state spectrum and
show that one pair of Fermi arcs appear in the surface of the
first Brillouin zone, which gives another evidence of the
emergence of the topological phase in our system.

\subsection{Magnetic monopoles at the Weyl points}
States at the Weyl points can be viewed as topological defects in
the momentum space and are extremely stable. For translationally
invariant system, infinitesimal transformation of the Weyl
Hamiltonian only shifts the Weyl points in energy or momentum, but
does not remove them from the energy spectrum. In this sense, the
Weyl points possess absolute stability, which hints on that the
Weyl points have topological protection and can be related to a
topological index. In fact, it is found that the topological
charges of the Weyl points are associated with quantized source of
Berry curvature. If we assume the ground state of the Weyl
Hamiltonian as $\psi(\textbf{q})$, the Berry curvature for this
Weyl node is expressed as
$\mathbf{\mathcal{F}}(\textbf{q})=i\langle\nabla_\textbf{q}\psi(\textbf{q})|\times|\nabla_\textbf{q}\psi(\textbf{q})\rangle$.
Then the topological charge of the Weyl point can be characterized
by the first Chern number
 \begin{equation}
 C=\frac{1}{2\pi}\oint_S \mathbf{\mathcal{F}}\cdot dS,
 \end{equation}
where the integration is perform on a closed surface $S$ in the
momentum space surrounding the Weyl point. In our system, with the
Weyl Hamiltonians (\ref{hw}), the Berry curvatures for the Weyl
points $W_{1,3}$ and $W_{2,4}$ are calculated as
$\mathbf{\mathcal{F}}_{1,4}=\textbf{q}/2|\textbf{q}|^3$ and
$\mathbf{\mathcal{F}}_{2,3}=-\textbf{q}/2|\textbf{q}|^3$,
respectively. In Fig. 2(b), we plot the Berry curvature around the
four Weyl points in our model. As one can see, a Weyl point can be
viewed as a magnetic monopole. The Berry curvature and the Berry
flux number are associated with the magnetic field and the
quantized magnetic charge of such monopole. From Fig. 2(b), one
can easily find that the Chern numbers (monopole charges) for the
Weyl points $W_{1,4}$ and $W_{2,3}$ are $C=1$ and $C=-1$, which
are exactly equal to the chiralities of the four Weyl points and
give the physical meaning of the chiralities. This result also
shows that our system has two pairs of Weyl points with opposite
monopole charges. This feature agrees well with the
fermion-doubling theorem which states that Weyl points must come
in pairs in a lattice system and the sum of their monopole charges
is zero \cite{FDT}.

 \begin{figure}[h]
\includegraphics[width=8cm,height=4.2cm]{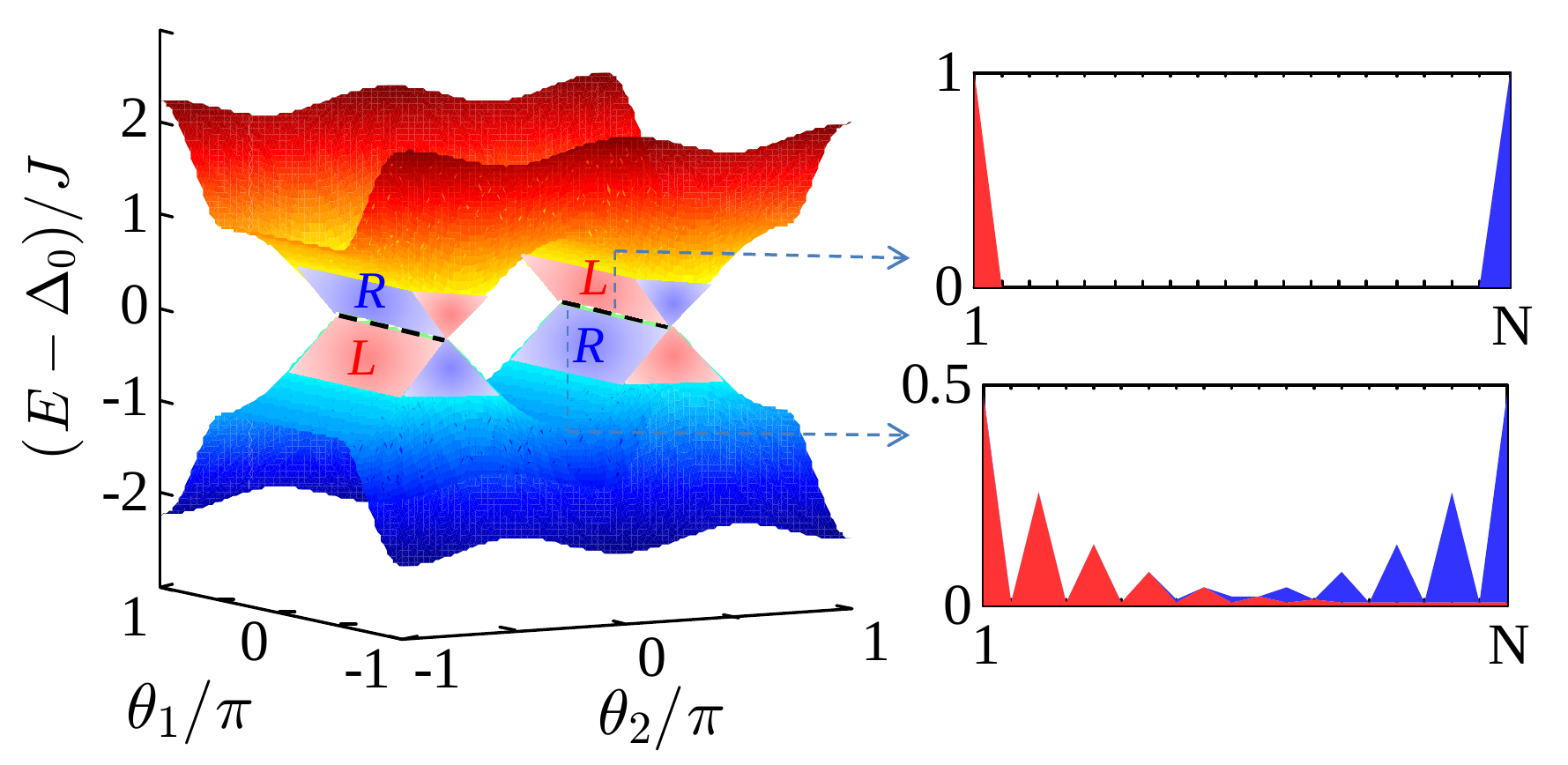}
\caption{Edge state spectrum in the surface Brillouin
zone. There are two sheets of edge states  denoted as $L$ and $R$
localized in the leftmost and the rightmost resonators. The Fermi
arcs are the intersection of the two sheets of edge states. The
density distributions of two Fermi arc states are plotted on the
right hand side. The lattice size is chosen as $N=20$ and other
parameters are the same as those in Fig. 2. }
\end{figure}

\subsection{Topologically protected surface states in the Fermi arc}
Another distinguishing topological feature for the Weyl semimetal
phase is the existence of topologically protected  surface states.
This feature is quite counter-intuitive because Weyl semimetal is
a gapless system, where the surface states are expected to mix
with the Bulk states. However, the Weyl semimetal system is
translationally invariant and the momentum is a good quantum
number. In this case, the surface states can be stabilized at the
momenta where there is no bulk state. Moreover, the surface states
at the Fermi energy in a Weyl semimetal form a Fermi arc, in
contrast to the usual Fermi surface that forms a closed loop. In
Fig. 3, we calculate the surface states of our model when the
circuit-QED lattice is under an open boundary condition. The
result shows that there are two sheets of edge states around zero
energy (the analog of Fermi energy), which are localized in the
leftmost and the rightmost resonators, respectively. The left and
the right edge states are associated with the surface states
localized on the two surfaces of our artificial 3D system. The
Fermi arcs appear as the intersection (dashed lines in Fig. 3) of
the two sheets of the edge states. To be more precise, there are
two Fermi arcs in our system, which appear as two lines that
connect the projections of the Weyl points $W_{1}$ ($W_{2}$) and
$W_{4}$ ($W_{3}$) on the surface of the Brillouin zone. More
specifically,
\begin{equation}
\begin{aligned}
\text{First Fermi arc: } (\theta_1,\theta_2)=(-\frac{\pi}{2}, \frac{\pi}{2})&\leftrightarrow(\frac{\pi}{2}, \frac{\pi}{2})\\ \nonumber
\text{Second Fermi arc: } (\theta_1,\theta_2)=(-\frac{\pi}{2}, -\frac{\pi}{2})&\leftrightarrow(\frac{\pi}{2},
-\frac{\pi}{2}). \nonumber
\end{aligned}
\end{equation}
This is because the edge states are ill-defined in the projection
of the Weyl points, where the gap is closed and there are bulk
states.  In Fig. 3, in order to verify this, we plot the density
distributions of two selected edge states, one of which is close
to the center of the Fermi arc and the other is near the
projection of the Weyl point. It turns out that the edge states
near the projection of the Weyl point penetrate more into the bulk
of the lattice; whereas the state near the center of the Fermi arc
is mainly located at the edge of the lattice. This result explains
why the Fermi arcs end at the projections of the Weyl points.
Although our system is bosonic by nature, one can find that the
spectrum of the surface state is the same as that of a fermionic
system and the zero energy in the spectrum is analog to the Fermi
energy. We will later show that the main features associated with
the surface states can be clearly observed in such photonic
systems.

\section{Detection of topological features in small lattice}
In this section, we will show that the essential topological
features of Weyl semimetals can be clearly measured using  a
cavity input-output process. The measurement can be performed by
driving the resonator lattice appropriately and detecting the
steady-state cavity output in the presence of finite resonator
dissipation. Intuitively, we may think that the introduction of
resonator dissipation could destroy the topological states.
However, our results show that dissipation is in fact beneficial
for measuring the topological features in Weyl semimetals.
Especially, we find that the monopole charges of the Weyl points
and the Fermi arcs can be unambiguously measured  in a circuit
with only four dissipative resonators. In other words, our scheme
does not require a large resonator lattice and is robust in
practical systems with finite  dissipation. This finding hence
greatly relaxes the experimental requirement and thus provides a
minimal system to observe the topological Weyl semimetal phase.

\subsection{Lattice-based cavity input-output process}
Different from fermionic systems, our model employs a bosonic
system where multiple photons can occupy one eigenstate at  the
same time. Inspired by this fact, one can excite the circuit-QED
lattice to occupy a particular energy eigenstate by applying
appropriate external microwave field driving at designated driving
frequency to change $\Delta_0$ and matching to the eigenenergy of
the whole lattice, and also by applying the driving on proper
resonator so that it has the maximal wave function overlap with
the eigenstate of the lattice. With this method, one can drive the
lattice to occupy the photonic edge states or bulk states,
depending on the choice of the driving frequency and the driven
lattice sites.

The above lattice-based method and the final steady state can be
concretely formalized as follows. Each resonator in the lattice is
firstly assumed to be driven by an external microwave pulse. In
the rotating frame with respect to the driving frequency
$\omega_d$, the driving Hamiltonian is
\begin{equation}
H_d=\sum_{n}(\Omega_{na}a^{+}_n+\Omega_{nb}b^{+}_n)+H.c,
\end{equation}
where $\Omega_{na,nb}$ are the driving amplitudes on the two
resonators in the $nth$ unit cell. When the  resonator dissipation
is taken into account, the final steady state of the resonator
lattice can be solved using a standard Lindblad master equation
for this system. In particular, the expectation value of the
resonator field $a_j$ in the steady state can be expressed as
\begin{equation}
\label{master}
\langle\dot{a_j}\rangle=-i\langle[a_j,H+H_d]\rangle+\kappa\sum_n\langle L[a_n]a_j\rangle,
\end{equation}
where the Lindblad term $L[a_n]a_j=a_n
a_ja^{+}_n-\{a^{+}_na_n,a_j\}/2$ and $\kappa$ is the resonator
decay rate.  To obtain a simple formula for the expectation values
of all the resonators in the lattice, we choose to work in the new
bases $\vec{a}=(\langle a_1\rangle,\langle b_1\rangle,...,\langle
a_n\rangle,\langle b_n\rangle)^T$ and
$\vec{\Omega}=(\Omega_{1a},\Omega_{1b},...,\Omega_{na},\Omega_{nb})^T$.
Based on the condition of the steady-state solution
\begin{equation}
\langle\dot{a_j}\rangle=0,
\end{equation}
we can get the expectation value of all the resonator fields in the steady state and further write them in a compact form
\begin{equation}
\label{steady}
\vec{a}=-(\Delta_0+T-i\frac{\kappa}{2})^{-1}\vec{\Omega},
\end{equation}
where the elements of matrix $T$ are defined by $T_{na,nb}=T_{nb,na}=J_1$, $T_{na,(n-1)b}=T_{(n-1)b,na}=J_2$, $T_{na(b),na(b)}=\pm J_e\cos(\theta_2)$.

In the above process, the external drive can be referred as the
process of injecting photons to the resonators, and thereby the
above scheme is naturally related to the cavity input-output
process. In our work, as we will demonstrate below, the
topological features of the photonic Weyl semimetal can be
observed by driving the resonator lattice to occupy the edge
states and measuring the resulted steady-state response. In
particular, we choose to excite the edge states localized in the
left edge of resonator lattice with the driving frequency tuned to
match the corresponding eigenenergy and the driving pulse applied
only to the leftmost resonator, as the left edge mode has the
maximal probability of populating the leftmost resonator. In this
case, the driving pulses have a form
$\vec{\Omega}=(\Omega_{1a},0,...,0,0)^T$, where $\Omega_{1a}$ is
the external driving amplitude. According to the cavity
input-output formula, the output photon field $a^{out}_1$ from the
resonator at the left edge is related to the input photon through
\cite{QObook}
\begin{equation}
a^{out}_1=a^{in}_1+\sqrt{\kappa}a_1,
\end{equation}
where the input field $a^{in}_1$ is related to the external driving by \cite{rmpclerk}
\begin{equation}
\sqrt{\kappa}a^{in}_j=i\Omega_{1a}.
\end{equation}
Combining this equation with Eq.~(\ref{steady}), the reflection coefficient from the left edge can be derived as
\begin{equation}
\label{rf}
r_L=\frac{\langle a^{out}_1\rangle}{\langle a^{in}_1\rangle}=1+i\kappa[(\Delta_0+T-i\frac{\kappa}{2})^{-1}]_{11}.
\end{equation}
In the following, we will demonstrate that this reflection
coefficient can be used to measure all the essential features  of
the photonic Weyl semimetal in our system.

\begin{figure}[h]
\includegraphics[width=8cm,height=8cm]{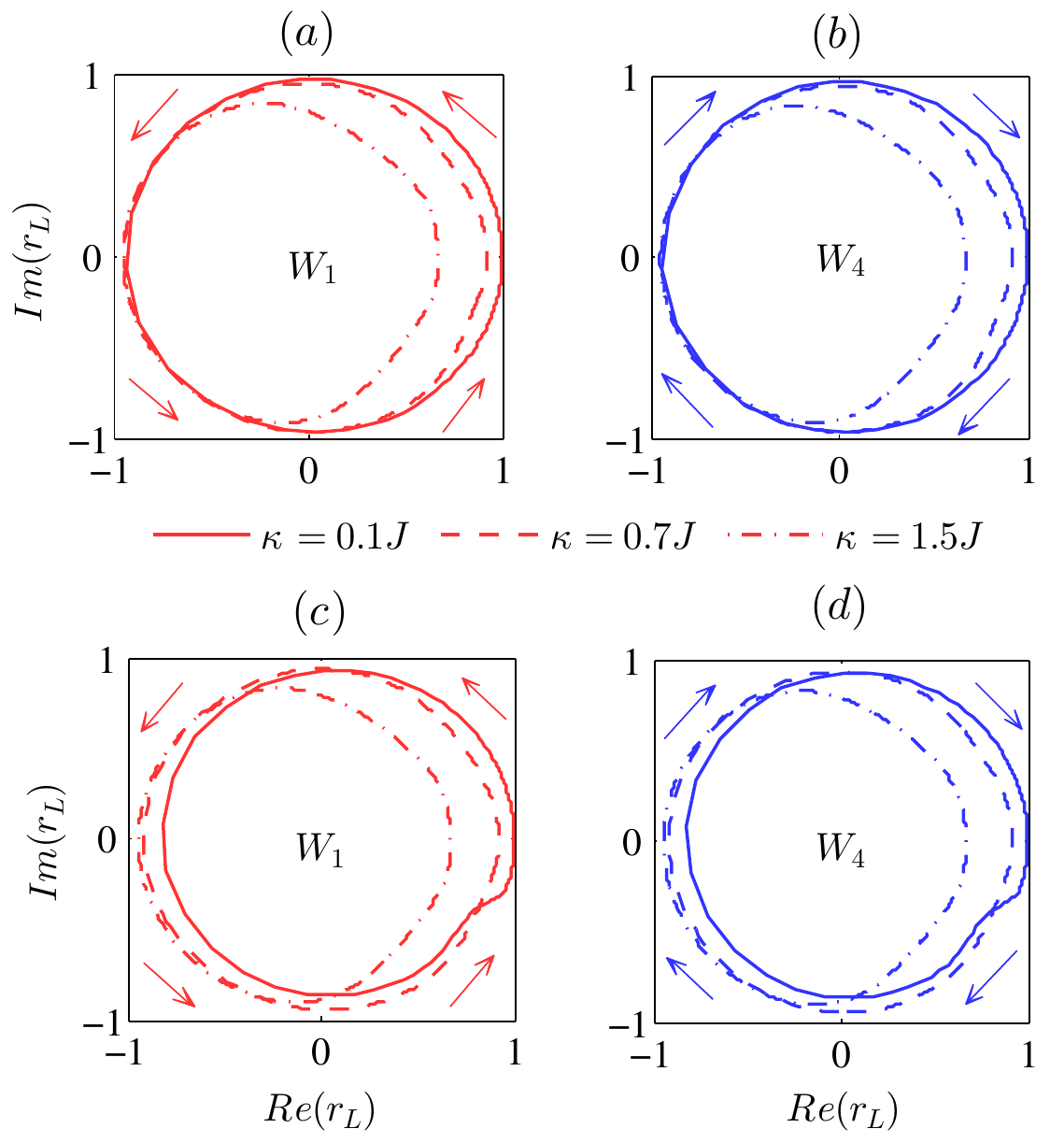}
\caption{The winding of the reflection coefficient phase as a
function of $\theta$ for the Weyl point (a) $W_1$ and (b) $W_4$,
with a lattice size $N=12$. The two cases are also shown in (c)
and (d), respectively, for a lattice size $N=4$. The detuning is
$\Delta_0=-0.1J$, and the resonator dissipation rate is $\kappa =
0.1J$ (solid line), $\kappa = 0.7J$ (dashed line), and
$\kappa=1.5J$ (dash-dotted line). The arrows $\circlearrowleft$
and $\circlearrowright$ signify the winding directions as
anticlockwise and clockwise (defined in right-handed coordinate
system), which stand for the positive and negative signs of the
corresponding winding numbers. }
\end{figure}

\subsection{Measuring monopole charges of the Weyl points}
As shown before, the monopole charge or the chirality of a Weyl
point is quantified by the first Chern number of the ground state
on a momentum sphere surrounding the Weyl point. Meanwhile, we
know that two-dimensional (2D) topological insulators with no
symmetry are characterized by the first Chern numbers defined in a
2D momentum space. In this sense, we can represent the 3D momentum
sphere surrounding the Weyl point with two periodic quasimomentums
and further use them to construct an artificial 2D momentum space.
In such space, the Hamiltonian around the Weyl point becomes
gapped and can have a topological insulator state. In this way,
the topological feature of the Weyl point is mapped into that of a
topological insulator. The resulted merit is that the measurement
of monopole charge of a Weyl point is transferred to the
measurement of the topological invariant of a photonic topological
insulator state, which avoids the experimental challenges in
directly measuring the monopole charge.

To map the 3D momentum sphere $S$ around the Weyl point
$(k_{xw},\theta_{1w},\theta_{2w})$ into an effective  2D momentum
space, we first employ a circle to represent the 2D momentum
surface around the Weyl point in the $(\theta_{1},\theta_{2})$
plane, with the Weyl point as the center of this circle. In this
case, the momenta around the Weyl point in the
$(\theta_{1},\theta_{2})$ plane can be written in the following
simple form
 \begin{equation}
 \begin{aligned}
 \theta_1&=\theta_{1w}+\theta_{r}\cos(\theta),\\
 \theta_2&=\theta_{2w}+\theta_{r}\sin(\theta),
 \end{aligned}
 \end{equation}
where $\theta_r$ is the radius of the circle and the parametric
angle $\theta$ is within the range $0$ and $2\pi$.  If the radius
$\theta_r$ is smaller than half of the separation between two Weyl
points, one can find that the 3D momentum sphere around each Weyl
point can be safely simplified to a 2D momentum space formed by
$(k_x,\theta)$, where $\theta$ plays the role of an artificial
momentum dimension. Under this mapping, the Hamiltonian
$h(\textbf{k}_\textbf{l})$ around each Weyl point becomes
$h(k_x,\theta)$, and the monopole charge of each Weyl point has a
standard form of a Chern number
  \begin{equation}
 C=\frac{1}{2\pi}\int\int dk_xd\theta \mathbf{\mathcal{F}}(k_x,\theta),
 \end{equation}
where $\mathbf{\mathcal{F}}(k_x,\theta)$ is the Berry curvature of
ground band defined in the 2D momentum  space
\{$k_x\in(0,\pi],\theta\in(0,2\pi]$\}. In fact, such Chern number
is nothing but the topological index characterizing the
topological phase implied in the mapped Hamiltonian
$h(k_x,\theta)$. Numerically, we have calculated the above Chern
numbers associated with the four Weyl points in our system. The
result shows that their values agree well with their monopole
charges (chiralities). So the energy spectrum corresponding to
each Hamiltonian $h(k_x,\theta)$ will become gapped and supports a
topological insulator state, with its topological invariant
exactly as the monopole charge of the Weyl point.

Some of the present authors have recently designed a strategy to
measure the photonic topological invairant based on a  cavity
input-output process \cite{Mei2015}. The method shows that when
the in-gap edge mode of the photonic topological insulator system
is driven by external input photons, the photonic topological
invairant is just the winding number of the reflection coefficient
phase varying with $\theta$, which can be tracked by the cavity
input-output process. It is straightforward to apply this method
here to measure the photonic topological invariant of the ground
state of the mapped Hamiltonian $h(k_x,\theta)$. To this end, we
use the lattice-based cavity input-output method presented above
to drive the in-gap left edge states. The reflection coefficients
from the leftmost resonator $r_L(\theta)$ formulated in Eq.
(\ref{rf}) are numerically calculated and the results are plotted
in Fig. 4, corresponding to the mapped Hamiltonians around the
Weyl points $W_1$ and $W_4$. In the plots, we also present the
winding directions of the phase of reflection coefficients when
$\theta$ is tuned from $0$ to $2\pi$. The results in Fig. 4(a-b)
is with respect to the case when the lattice size is $12$. The
results show that the winding numbers corresponding to the Weyl
points $W_1$ and $W_4$ are $1$ and $-1$, which give the
measurement outcomes of the monopole charge of these two Weyl
points. Furthermore, we also calculate the case in Fig. 4(c-d)
when the lattice size is reduced to $4$. Remarkably, the winding
numbers remain the same as the case with a larger lattice. In all
cases, we also take into account  the influence of the cavity
dissipation. It is found that the winding number is clearly
observable if the cavity dissipation rate is smaller than the
energy gap of $2J$. The reason is that the frequency of input
photons (analog of Fermi level) will remain in the energy gap.
Then the corresponding topological invariant will remain the same.
In this sense, the above measurement is very robust to the
fluctuation of the frequency of input photons. The reflection
coefficient phase can be measured using the standard homodyne
detection techonology. Therefore, the monopole charges of Weyl
points could be clearly measured even in a circuit with only four
resonators.

 \begin{figure}[h]
\includegraphics[width=8.5cm,height=10cm]{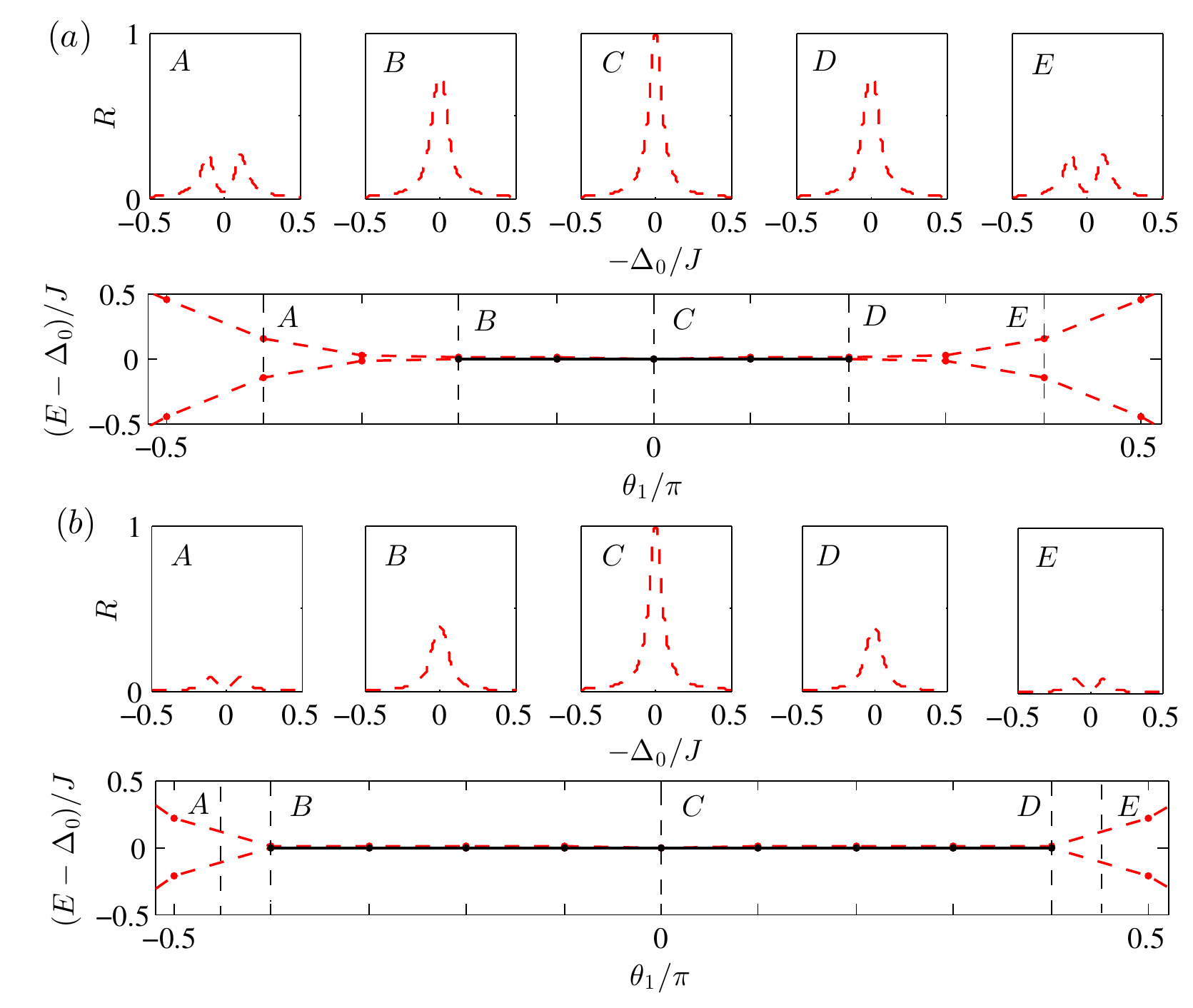}
\caption{(a) The intersection (dashed line) of edge states between
the projection of the Weyl  point $W_1$ and $W_4$ is plotted in
the lower panel for a lattice size of $N=4$. The reflection
spectra around zero energy with respect to five particular
$\theta_1$ values are plotted in the upper panel. The measured
Fermi arc based on such reflection spectra is also shown (the
solid line). The same results are shown in (b) for a lattice size
of $N=12$. The resonator dissipation rate is chosen as
$\kappa=0.1J$. }
\end{figure}

\subsection{Measuring the Fermi arcs}
In the last section, we have shown that Fermi arcs break off at
the projections of the Weyl points on  the surface Brillouin zone.
In particular, we study in detail how to measure the first Fermi
arc line that ends at the projection of the Weyl points $W_1$ and
$W_4$, labelled as $(\theta_1=\pm0.5\pi,\theta_2=0.5\pi)$. Such
Fermi arc can be measured by probing the position of zero energy
left or right edge modes. That is because Fermi arcs are the
intersection lines of the left and right edge states at zero
energy. In fact, we only need to probe the position of zero energy
edge modes along $\theta_1$ axis when the systematic parameter is
tuned to fix $\theta_2=0.5\pi$. With the lattice-based cavity
input-output process, the position of zero energy edge modes in
the $\theta_1$ direction can be measured based on externally
driving the edge resonators and then scanning the driven frequency
around zero energy to extract the reflection spectra
$R=|r_L(\Delta_0)|^2$. Each zero energy peak in such reflection
spectra would give the position of zero energy edge mode. The
reason is because only zero energy edge mode is a resonant
eigenmode of the photonic lattice in the above external driven.
Such mode has the maximal probability to populate the lattice,
while the other non-resonant mode would quickly decay into the
vacuum state.

In our case, we employ a four coupled resonator lattice and the
first Fermi arc line is numerically calculated in  the down panel
of Fig. 5(a). It is found that the Fermi arc ends at the points
with $\theta_{1c}=\pm0.2\pi$ instead of $\pm0.5\pi$. When
$\theta_1$ is bigger than $\theta_{1c}$, an energy gap appears and
the intersection Fermi arc line disappears. The reason for this
mismatch is due to finite lattice size effect. To measure this
Fermi arc, we choose to drive the leftmost resonator to excite the
left edge state. In particular, we have calculated the reflection
spectra around zero energy for five $\theta_1$ values in the upper
panel of Fig. 5(a), including three inside and two outside the
Fermi arc. The results show that the reflection spectra for
$\theta_1$ values inside the Fermi arc has a peak at the zero
energy, which means that this mode is the left Fermi arc edge
state. In this way, the whole Fermi arc shape is mapped out and
then the Fermi arc end points are measured with
$\theta_{1c}=\pm0.2\pi$. Similarly, we also calculate the  Fermi
arc line and the reflection spectra for a lattice size of $12$ in
Fig. 5(b). It turns out that the Fermi arc ends at the points with
$\theta_{1c}=\pm0.4\pi$. The measured Fermi arc ending points
derived from the reflection spectra also agree well with this
point. In both case, we also plot the reflection spectra for
$\theta_1$ values outside the Fermi arc. It turns out that there
appears two symmetric peaks around zero energy, which agree with
the fact that an energy gap will appear when $\theta_1$ is outside
the Fermi arc. Moreover, one also can find that the heights of the
peaks in the reflection spectra become very low when $\theta_1$ is
closer to $\pm0.5\pi$. As we have shown in Section IV.B, that is
because edge states near the projection of the Weyl point
penetrate more into the bulk and the probability occupying the
leftmost or rightmost resonator will be very small.

We have summarized the finite lattice size effect on the measured
Fermi arc ending points in Table 1.  The results show that the
measured Fermi arc ending points $\theta_{1c}$ would be closer to
the projections of Weyl points $\theta_{1}=\pm0.5\pi$ when the
lattice size is increased to $20$. However, the cost of getting
more closer is very high and it requires a larger lattice size.
That is because the Fermi arc edge state is ill defined at the
projection of the Weyl point. So, it is not necessary to observe
the behavior of fermi arc approaching the projection of Weyl
point. In this sense, we conclude that the Fermi arcs can be
measured in a circuit with only four dissipative resonators, which
thus provides us a minimal platform to observe the topological
Fermi arcs.

\begin{table}[tbp]
\centering
\caption{Finite lattice size effect on the measured Fermi arc ending points.}
\begin{tabular}{ccccccc}
\hline \hline
Parameters & & &Values & & & \\ \hline
Lattice size $N$ &4 &6 &8 &12 &20 &36 \\    \hline
$\theta_{1c}$ &$\pm0.2\pi$ &$\pm0.3\pi$ &$\pm0.35\pi$  &$\pm0.4\pi$ &$\pm0.45\pi$ &$\pm0.48\pi$ \\     \hline
\end{tabular}
\end{table}

\section{Summary}
In summary, we have presented a minimal circuit-QED lattice model
to mimic the topological features of a Weyl semimetal. Moreover,
we have shown that both the topological charges associated with
the Weyl points and the Fermi arcs can be clearly measured in a
small circuit with only four dissipative resonators. Considering
recent advances in the technology for controlling arrays of
superconducting resonators and qubits, our scheme can be realized
in practical systems. This work hence can  stimulate further
experimental and theoretical interests on quantum simulation of
topological phases with small artificial lattice systems.

\begin{acknowledgements}
This work was supported by the National Basic Research Program of China (Grant No. 2012CB821305), the NSF of China (Grants No. 11474153, No. 11374375 and No. 11574405), and the PCSIRT (Grant No. IRT1243). L. T. was supported by the National Science Foundation under Award No. NSF-DMR-0956064. Z. X. was supported by NFRPC (Grant No. 2013CB921804). D. Z. was supported by the NSF of Guangdong Province (Grant No. 2016A030313436), the FDYT (Grant No. 2015KQNCX023) and the Stratup Foundation of SCNU.
\end{acknowledgements}

\end{document}